\documentclass[aps,prl,twocolumn,superscriptaddress,showpacs]{revtex4-1}
\usepackage{graphics}
\usepackage{graphicx}
\usepackage{ulem}
\usepackage{amsmath}
\ifx\pdftexversion\undefined
\usepackage[dvips]{hyperref}
\else
\usepackage{hyperref}
\fi
\hypersetup{
  colorlinks = true, linkcolor = blue
}

\begin{document}

\title{Momentum-resolved observation of thermal and quantum depletion in a Bose gas}

\author{R. Chang}
\affiliation{Laboratoire Charles Fabry, Institut d'Optique, CNRS, Univ. Paris Sud, 2 Avenue Augustin Fresnel 91127 PALAISEAU cedex, France}
\author{Q. Bouton}
\affiliation{Laboratoire Charles Fabry, Institut d'Optique, CNRS, Univ. Paris Sud, 2 Avenue Augustin Fresnel 91127 PALAISEAU cedex, France}
\author{H. Cayla}
\affiliation{Laboratoire Charles Fabry, Institut d'Optique, CNRS, Univ. Paris Sud, 2 Avenue Augustin Fresnel 91127 PALAISEAU cedex, France}
\author{C. Qu}
\affiliation{INO-CNR BEC Center and Dipartimento di Fisica, Università di Trento, 38123 Povo, Italy}
\author{A. Aspect}
\affiliation{Laboratoire Charles Fabry, Institut d'Optique, CNRS, Univ. Paris Sud, 2 Avenue Augustin Fresnel 91127 PALAISEAU cedex, France}
\author{C. I. Westbrook}
\affiliation{Laboratoire Charles Fabry, Institut d'Optique, CNRS, Univ. Paris Sud, 2 Avenue Augustin Fresnel 91127 PALAISEAU cedex, France}
\author{D. Cl\'ement}
\email[To whom correspondence should be adressed: ]{david.clement@institutoptique.fr}
\affiliation{Laboratoire Charles Fabry, Institut d'Optique, CNRS, Univ. Paris Sud, 2 Avenue Augustin Fresnel 91127 PALAISEAU cedex, France}

\begin{abstract}
We report on the single-atom-resolved measurement of the distribution of momenta $\hbar k$ in a weakly-interacting Bose gas after a 330 ms time-of-flight.  We investigate it for various temperatures and clearly separate two contributions to the depletion of the condensate by their $k$-dependence. The first one is the thermal depletion. The second contribution falls off as $k^{-4}$, and its magnitude increases with the in-trap condensate density as predicted by the Bogoliubov theory at zero temperature. These observations suggest associating it with the quantum depletion. How this contribution can survive the expansion of the released interacting condensate is an intriguing open question.
\end{abstract}


\maketitle


In quantum systems, intriguing many-body phenomena emerge from the interplay between quantum fluctuations and interactions. Quantum depletion is an emblematic example of such an effect, occurring in one of the simplest many-body systems: a gas of interacting bosons at zero temperature. In the absence of interactions, the ground state corresponds to all particles occupying the same single-particle state. Taking into account inter-particle repulsive interactions at the mean-field level leads to a similar solution where all particles are condensed in the same one-particle state whose shape is determined by the trapping potential and interactions. In a beyond mean-field approach, which can be interpreted as taking into account quantum fluctuations and two-body interactions, the description is dramatically different. The many-body ground state consists of several components: a macroscopically occupied single-particle state, the condensate, and a population of single-particle states different from the condensate, the depletion. 

This many-body description applies to diverse bosonic systems such as superfluid Helium \cite{Sokol1995}, exciton-polaritons \cite{Utsunomiya2008} and degenerate Bose gases \cite{Dalfovo1999}; it has also found analogies in phenomena such as Hawking radiation from a black-hole \cite{Hawking1975} and spontaneous parametric down conversion in optics \cite{Yuen1976}. The fraction of atoms not in the condensate at zero temperature, the quantum depletion, increases with the strength of inter-particle interactions and with the density, rising up to 90\% in liquid $^4$He \cite{Sokol1995}. In ultracold gases, where the density is significantly smaller, the quantum depletion usually represents a small fraction (less than 1\%) of the total population. At non-zero temperature there is an additional contribution to the population of single-particle states above the condensate, originating from the presence of thermal fluctuations.  

For weakly interacting systems, Bogoliubov theory describes quantum and thermal contributions to the condensate depletion \cite{Bogoliubov1947, Lee1957}. This approach shows that the elementary, low-energy excitations are collective quasi-particle (phonon) modes, as has been verified in experimental studies with liquid $^4$He \cite{Donelli1981}, degenerate quantum gases \cite{Ozeri2005} and exciton-polaritons \cite{Utsunomiya2008}. At zero temperature, the many-body ground state is defined as a vacuum of these quasi-particle modes. When projected onto a basis of single-particle states with momentum $\hbar k$, this many-body ground state exhibits a distribution $n(k)$, which scales as $k^{-4}$ at large $k$. These power law tails do not exist in mean-field descriptions, for which the momentum distribution has a finite extent. At non zero temperature, the contribution to $n(k)$ induced by thermal fluctuations decays exponentially for energies larger than the temperature. Previous experiments with atomic gases \cite{Kohl2004, Xu2006} have observed the total depletion of the condensate after a time-of-flight expansion, but could not distinguish between the thermal and quantum contributions. 

\begin{figure}[ht!]
\includegraphics[width=\columnwidth]{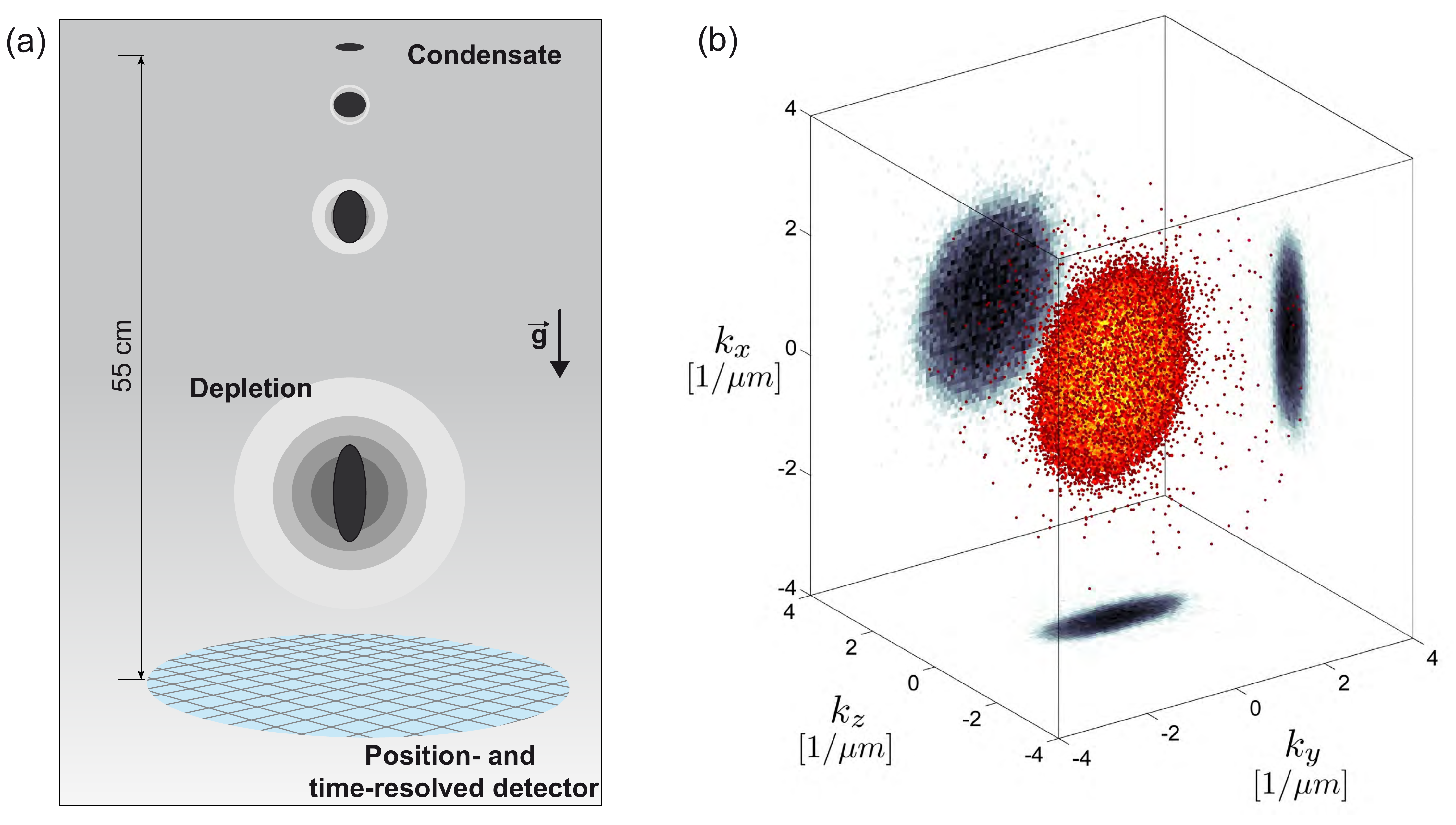}
\caption{{\bf (a)} Sketch of cloud expansion and detection by a micro-channel plate detector, yielding the 3D asymptotic momentum distribution (far-field regime). The initially cigar-shaped Helium condensate (black) undergoes anisotropic expansion, inverting its aspect ratio. Quantum depletion and/or thermally excited atoms (grays) populate momentum states beyond those associated with the condensate, and are expected to have a spherical symmetry. {\bf (b)} Measured 3D distribution of atoms $n_\infty({\bf k})$ after a 330 ms time-of-flight.  The central dense part corresponds to the condensate while the isolated dots are excited particles outside of the condensate wavefunction. Also shown are the 2D projections, highlighting the condensate anisotropy. \label{Fig1}}
\end{figure}

\begin{figure}[ht!]
\includegraphics[width=\columnwidth]{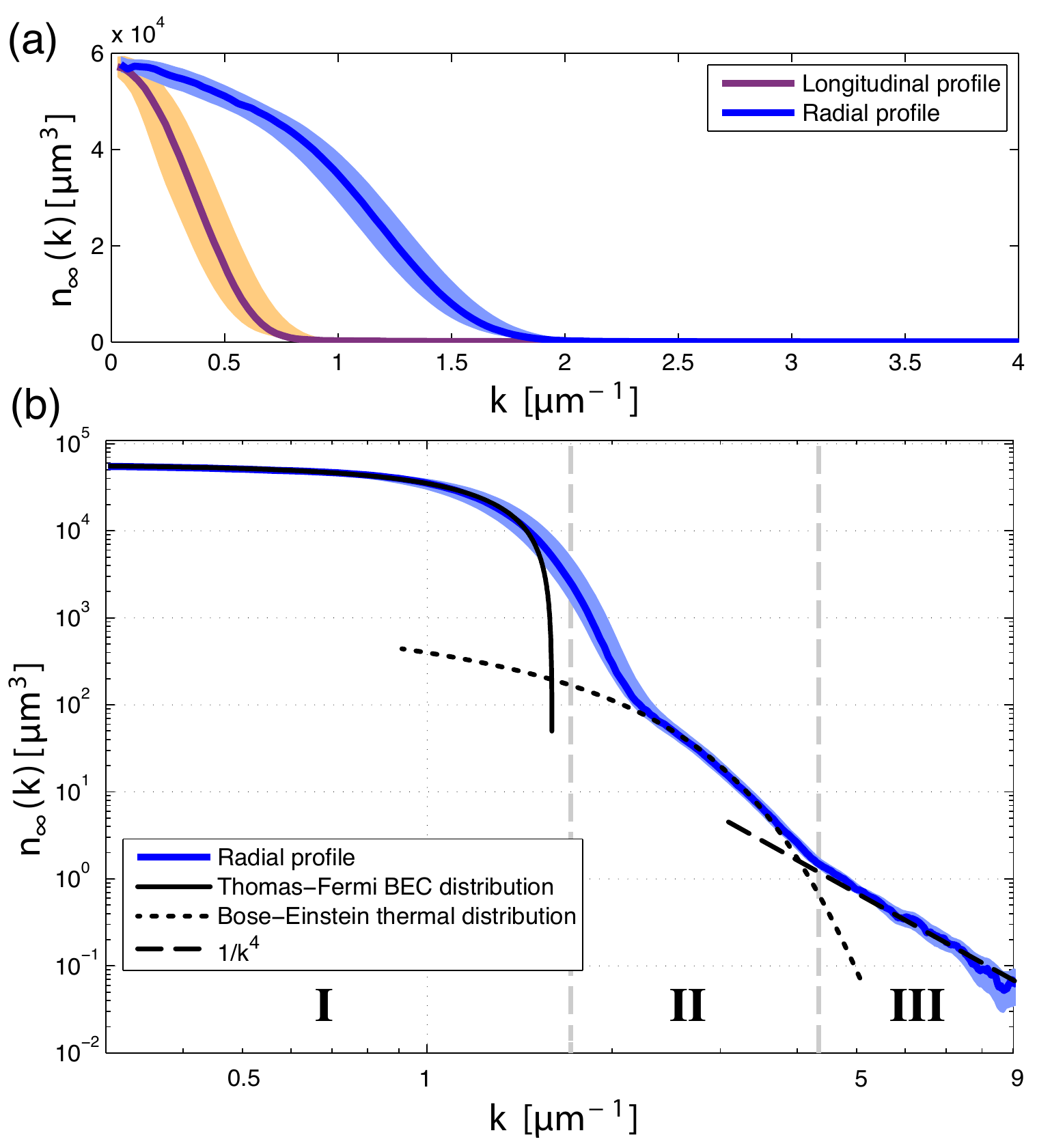}
\caption{1D momentum profiles obtained from cuts of the 3D data $n_\infty({\bf k})$. {\bf (a)} Profiles along the radial (blue) and longitudinal (purple) directions. In linear scale, the tails are not visible {\bf (b)} Log-log scale plot of the radial profile. The solid line is the scaling solution for the condensate in the Thomas-Fermi approximation (region \textbf{I}: $k<1.7\,\,\mu m^{-1}$); the dotted line is a Bose distribution fitting the thermal wings (region \textbf{II}: $1.7<k<4.3\,\,\mu m^{-1}$); the dashed line is a $k^{-4}$ power-law fitting the high-momentum tails (region \textbf{III}: $k>4.3\,\,\mu m^{-1}$). Solid lines are a smoothed average of the density data, and the lightly shaded band is the running standard deviation. The dark count rate corresponds to a level less than $\sim 10^{-2} \mu$m$^3$, which is one order of magnitude below the lowest data point. The plotted range of momenta is limited by the physical size of the detector. \label{Fig2}}
\end{figure}

In this letter, we report on the observation of momentum-space signatures of thermal and quantum depletion in a gas of interacting bosons. We investigate, for various temperatures and atomic densities, the three-dimensional atomic distribution after a long time-of-flight (see Fig.~\ref{Fig1}), {\it i.e.}, the  asymptotic momentum distribution. Three components can be identified (see Fig.~\ref{Fig2}):  the condensate ({\bf I}), the thermal depletion ({\bf II}) and a tail decaying as $k^{-4}$ and increasing with the in-trap condensate density ({\bf III}). This suggests associating region {\bf III} with the quantum depletion, but with two caveats. Firstly, $k^{-4}$-tails originated from contact interaction were observed to vanish during the expansion of interacting fermions \cite{Stewart2010}. Recent theoretical work investigating interacting bosons predicts that the $k^{-4}$-tails adiabatically decrease with the condensate density during the expansion \cite{Qu2016}. Secondly, the magnitude of the $k^{-4}$-tails we measure is larger than the in-trap prediction of the Bogoliubov theory, by a factor of about 6. Our identification of the $k^{-4}$-tail with the quantum depletion thus demands that there exists either a non-adiabatic process in the expansion, decoupling the in-trap $k^{-4}$ component from the expanding condensate, or an interaction-induced effect beyond the mean-field description of the expansion, leading to 1/$k^{4}$ tails.

Our experiment is performed with a Bose-Einstein condensate of metastable Helium-4 atoms ($^4$He$^*$). Cigar-shaped condensates, of typically $N=2\times 10^5$ $^4$He$^*$ atoms in the polarized $2^3S_1$, $m_{J}=+1$ state are produced in an optical dipole trap with trapping frequencies $\omega/2\pi= (438, 419, 89)$ Hz \cite{Chang2014, Bouton2015}. After abruptly turning off the optical trap (in less than $2~\mu s$) we detect the gas with a micro-channel plate (MCP) \cite{Schellekens2005, Nogrette2015}, after a 55 cm free-fall corresponding to a time-of-flight (TOF) of $\sim 330$ ms. A radio-frequency (RF) ramp transfers a fraction of the polarized $m_{J}=+1$ atoms to the non-magnetic $m_{J}=0$ state after 2 ms of TOF (see \cite{SuppMaterial} for details). The presence of a magnetic gradient after the RF ramp ensures that only $m_{J}=0$ atoms fall onto the MCP and can be detected. The MCP allows us to detect $^4$He$^*$ atoms individually and to record the two-dimensional (2D) position and the arrival time of each atom in the plane of the detector. The arrival time of each atom directly translates into a vertical coordinate, so that we reconstruct the complete 3D atoms distribution, in contrast with usual optical imaging which yields a 2D column-integrated density. Another advantage of a MCP operated in counting mode is its extremely low dark count rate. Here it allows us to investigate the atomic density over more than 5 decades (see below). 

The detector is placed 55 cm below the trapped cloud (see Fig.~\ref{Fig1}(a)), so that after switching off the trap, the observation is made in the far-field regime where finite-size effects of the source can be safely neglected. In the free-falling frame of reference, we thus identify the position ${\bf r}$ of a detected atom (with respect to the cloud center) with a momentum component ${\bf k}=m{\bf r}/\hbar \bar{t}$, where $\bar{t}=330$ ms is the time-of-flight of the cloud center \cite{SuppMaterial}. This yields the asymptotic momentum distribution $n_\infty({\bf k})$ obtained from the density distribution of the expanding cloud $n({\bf r},\bar{t})$, according to the ballistic relationship
\begin{equation}
n_\infty({\bf k}) = (\hbar \bar{t}/m)^3 \ n({\bf r},\bar{t})
\label{Eq:asymp}
\end{equation}
The distribution $n_\infty({\bf k})$ should not be confused with the in-trap momentum distribution $n({\bf k})$, since the initial phase of the expansion is affected by inter-atomic interactions. Nevertheless, as we argue below, the high-momentum tails of $n_\infty({\bf k})$ provide interesting information on the in-trap momentum distribution $n({\bf k})$.

An example of $n_{\infty}({\bf k})$ is shown in Fig.~\ref{Fig1}(b). The main component is the pancake-shaped distribution expected for the cigar-shaped condensate according to the mean-field description of the expansion \cite{Castin1996, Kagan1996}. We also distinguish momentum components beyond those of the condensate, with a much lower density and an isotropic distribution \cite{SuppMaterial}. From 3D distributions $n_{\infty}({\bf k})$, we obtain radial and longitudinal profiles as shown in Fig.~\ref{Fig2} \cite{SuppMaterial}. The resulting profiles exhibit three distinct regions, as illustrated in Fig.~\ref{Fig2}(b).


Firstly, the observed distributions are dominated by the high-density condensate (region {\bf{I}}: $k\equiv\vert{\bf k}\vert <1.7$ $\mu m^{-1}$). The initial expansion of the condensate is driven by inter-particle interactions, resulting in an asymptotic distribution different from the in-trap momentum distribution. This dynamics is fully captured by a mean-field treatment, the scaling solution \cite{Castin1996, Kagan1996} calculated in the Thomas Fermi approximation, as shown in Fig.~\ref{Fig2}(b). We have checked that a numerical solution of the time-dependent Gross-Pitaevskii equation for the pure condensate,  beyond the Thomas Fermi approximation, yields negligible corrections \cite{SuppMaterial}.  

Beyond the anisotropic mean-field distribution we observe high-momentum tails ($k>1.7\,\,\mu m^{-1}$) with spherical symmetry \cite{SuppMaterial}: we attribute these components (regions \textbf{II} and \textbf{III} in Fig.~\ref{Fig2}(b)) to the depletion of the condensate. The {\it isotropic} character of the atomic distribution in the tails indicates that the {\it anisotropic} mean-field potential describing the interactions in the condensate does not play a significant role in the expansion of particles belonging to regions \textbf{II} and \textbf{III}. Thus we assume in the following that the high-momentum components of the tails, corresponding to a kinetic energy larger than the chemical potential of the condensate, quickly escape the condensate and are not affected by the mean-field potential during the expansion. The tails are visible over three decades in density, allowing us to perform a detailed study of their momentum dependence. We observe two distinct regions: a middle region without a well-defined power-law variation (region {\bf{II}}), and a high-momentum region with density varying as $k^{-\alpha}$ (region {\bf{III}}). A power-law fit to the data in region {\bf III} yields $\alpha=4.2(2)$.


A quantitative description of the condensate depletion close to zero temperature is provided by Bogoliubov's microscopic theory, which yields a beyond mean-field model taking into account quantum and thermal excitations \cite{Bogoliubov1947, Lee1957}. In the many-body Hamiltonian, this approximated approach retains only quadratic terms in the particle operators $a_{k}$, where $a_{k}$ is the operator annihilating a particle with momentum $\hbar k$. The simplified Hamiltonian can then be diagonalized by introducing the quasi-particle operators $b_{k}$ defined by the Bogoliubov transformation $b_{k}=u_{k} a_{k} + v_{-k} a^\dagger_{-k}$ \cite{Bogoliubov1947, Lee1957}. 

At zero temperature, the many-body ground state is the vacuum of quasi-particles, defined as $\langle  b^\dagger_{k} b_{k} \rangle=0$ for any $k\neq0$. It corresponds to a non-zero population of excited single-particle states, $\langle a^\dagger_{k} a_{k} \rangle=|v_{k}|^2$ for $k\neq0$. This is the quantum depletion of the Bose condensate, which has no classical analog, and can be seen as arising from the interplay of Bose statistics and interactions. At non-zero temperature, the particle occupation number of non-zero momentum $k$ can be expressed analytically in terms of the quasi-particle occupation number:
\begin{equation}
n(k)=\langle a^\dagger_{k} a_{k} \rangle=(|u_{k}|^2 + |v_{k}|^2) \langle b^\dagger_{k} b_{k} \rangle +|v_{k}|^2\label{Eq:Nk}
\end{equation}
with the occupation number of quasi-particles given by a Bose-Einstein distribution,
\begin{equation}
\langle b^\dagger_k b_k \rangle=\frac{1}{\exp (\epsilon(k)/k_BT)-1}
\end{equation}
where $\epsilon(k)$ is the Bogoliubov dispersion relation, $k_B$ is Boltzmann's constant and $T$ is the temperature. The interpretation of 
Eq.~\ref{Eq:Nk} is clear: the first term, proportional to $\langle b^\dagger_{k} b_{k} \rangle$, represents the thermal depletion; the second term is associated with the quantum depletion.

\begin{figure*}[ht!] 
\includegraphics[width=2\columnwidth]{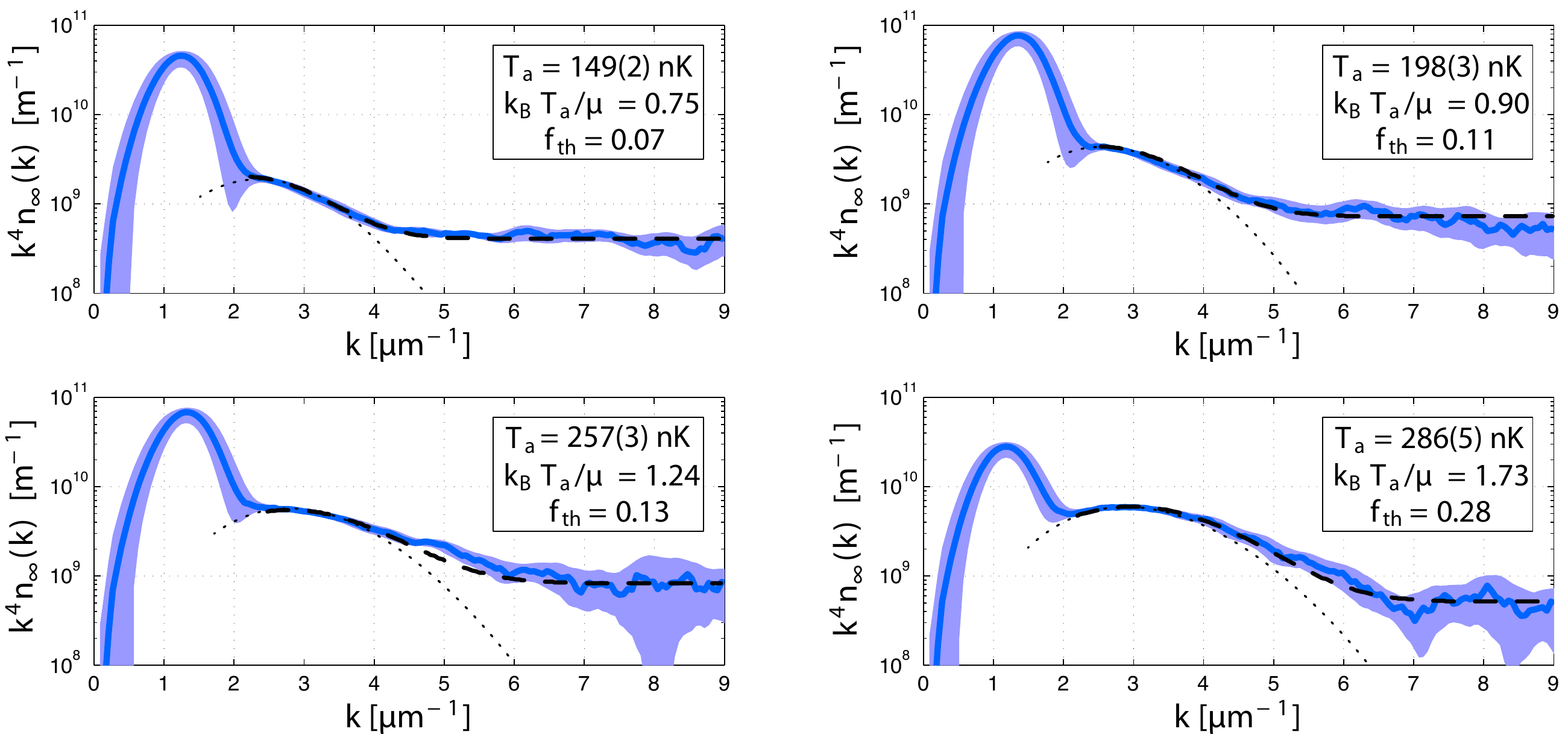}
\caption{Measured $k^4 n_{\infty}(k)$ plotted as a function of $k$ and at various temperatures. Solid blue lines are a smoothed average of the density data, and the lightly shaded blue band shows the running standard deviation. The dashed line is a fit using Eq.~\ref{Eq:Fit_tails} which involves two terms: the thermal depletion (also shown as a dotted line) and the quantum depletion revealed by the $k^{-4}$ scaling at large $k$. The fitting procedure yields the temperature $T_{a}$, the thermal atom number $N_{th}$ and the asymptotic constant $\mathcal{C_{\infty}}$. Noted in each subplot are $T_{a}$, the ratio of the thermal energy and the chemical potential $k_B T_a/\mu$ and $f_{th}=N_{th}/N$.
\label{Fig3}}
\end{figure*} 

The particle occupation number $|v_{k}|^2$ corresponding to the quantum depletion varies as $k^{-1}$  for $k\xi \ll 1$, and as $k^{-4}$ for $k\xi \gg1$, where $\xi$ is the healing length of the condensate. The small $k$ behavior is related to the Heisenberg inequality associated with the particle and the density operators \cite{Pitaevskii1991}. The large $k$ behavior arises due to the two-body contact interaction and is related to Tan's contact constant, a universal quantity that connects contact interactions to the thermodynamics of a many-body system \cite{Tan2008, Combescot2009}. By contrast, the depletion associated with the thermal excitations varies differently with $k$ due to the additional term $\langle b^\dagger_k b_k \rangle$. In particular, it decays exponentially for $k \lambda_{dB} \gg1$, where $\lambda_{dB}=\sqrt{h^2/2 \pi m k_{B} T}$ is the de Broglie wavelength. These differences provide a means to unambiguously distinguish the quantum depletion from the thermal depletion.

The presence of an inhomogeneous trap does not modify the prediction for the condensate depletion at momenta large compared to the inverse system size $1/R\simeq 0.08 \ \mu$m$^{-1}$, where the results of Bogoliubov theory can be averaged using the local density approximation (LDA). For a harmonically trapped gas, a LDA calculation keeps the large-momentum  $k^{-4}$ scaling of the homogeneous model \cite{SuppMaterial}. On the other hand, the thermal depletion distribution in a harmonic trap approaches a polylog function at high temperature \cite{Dalfovo1999}. 

To identify the contributions in regions {\bf II} and {\bf III}, we have investigated the tails of the measured distribution $n_\infty(k)$ as a function of temperature. Fig.~\ref{Fig3} presents the radial distributions $k^4 \times n_{\infty}(k)$ for clouds subjected to a controlled heating sequence \cite{SuppMaterial}. Assuming the LDA average of Eq.~\ref{Eq:Nk}, we fit the tails in Fig.~\ref{Fig3} ($k>2 \mu$m$^{-1}$)   with the function $k^4 \times n_{\text{fit}}(k)$ where
 
\begin{equation}
n_{\text{fit}}(k)= \frac{N_{th}  \ g_{3/2}[\exp(- k^2 \lambda_{dB}^2/4 \pi)]}{1.202 \ (2 \pi/\lambda_{dB})^3} +\frac{\mathcal{C_{\infty}}}{(2 \pi)^3k^4}\label{Eq:Fit_tails} 
\end{equation} 
with $T_{a}$ (via $\lambda_{dB}=h/\sqrt{2 \pi m k_{B} T_{a}}$), $N_{th}$ and $\mathcal{C_{\infty}}$ are fitting parameters. The first term in Eq.~\ref{Eq:Fit_tails} is the polylog function describing a thermal component with an atom number $N_{th}$ and an apparent temperature $T_{a}$ \cite{NoteTempHartree}. The second term corresponds to a distribution decaying as $1/k^4$. The function $n_{\text{fit}}(k)$ is an excellent fit to the experimental profiles (see Fig.~\ref{Fig3}). As the gas is heated, the temperature $T_{a}$ and the thermal fraction $f_{th}=N_{th}/N$ increase. The variation of $f_{th}$ with $T_{a}/T_{c}$ ($T_{c}$ being the critical temperature of condensation) is in excellent agreement with the semi-classical prediction \cite{Dalfovo1999}, confirming our identification of region {\bf II} with the thermal depletion. Although they represent less than $\sim 0.5$ \% of the total atom number, the $k^{-4}$ tails are clearly visible beyond the thermal component (see Fig.~\ref{Fig3}), and thus associated with a zero-temperature effect \cite{NoteThermalContact}. In the weakly interacting regime we investigate, condensate lifetimes are on the order of seconds. We have measured identical $k^{-4}$-tails when holding the atoms for an extra second in the trap, showing that the gas is at equilibrium before the release.

The presence of $k^{-4}$-tails in a cloud released from a trap was previously reported in strongly interacting Fermi gases \cite{Stewart2010} but was not found with bosons \cite{Makotyn2014}. The observation of the $k^{-4}$-tails in a Fermi gas required ramping the interaction strength to zero before the expansion \cite{Stewart2010}, on a time scale shorter than that associated with many-body effects. Recent theoretical work concluded that the $k^{-4}$-tails should adiabatically vanish during the expansion of a Bose gas when the strength of interaction is kept constant \cite{Qu2016}. These considerations indicate that in order to associate the observed  $k^{-4}$-tails in $n_\infty(k)$ with the quantum depletion in the trapped cloud, we must invoke a non-adiabatic process. Since the scattering length in the $m_{J}=0$ state is expected to be smaller than that in the $m_{J}=+1$ state \cite{SuppMaterial, Leo2001}, a non-adiabatic transfer between these two states at the optical trap turnoff might explain our observation, but we have not yet found any evidence of this possibility in the experiment. On the other hand, there is no many-body treatment of the expansion of an interacting Bose gas. We thus cannot exclude the possibility that the tails result from a modification of the beyond mean-field momentum distribution during the time-of-flight dynamics.

\begin{figure}[ht!]
\begin{center}
\includegraphics[width=\columnwidth]{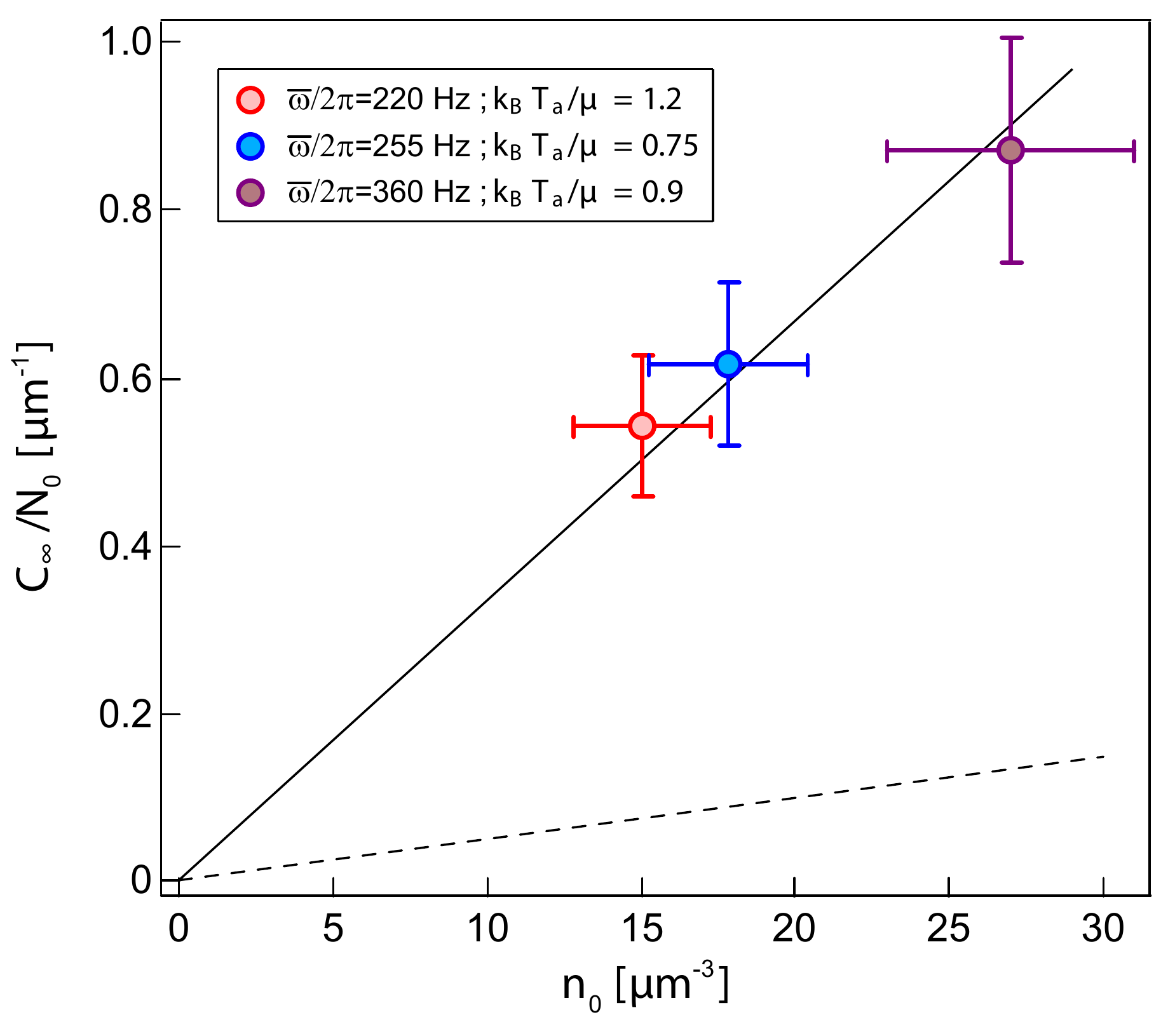}
\caption{Contact constant $\mathcal{C_{\infty}}/N_{0}$ per condensed particle plotted as a function of the condensate density $n_{0}$. The geometric trapping frequency ${\bar \omega}/2\pi$ and the ratio $k_BT_a/\mu$ are indicated. The dashed line is the Bogoliubov prediction in the LDA, $\mathcal{C}_{\text{LDA}}$ (see text), and the solid line is $6.5 \times \mathcal{C}_{\text{LDA}}$. \label{Fig4}}
\end{center}
\end{figure}

In order to further investigate the origin of the  $k^{-4}$-tails, we have studied their dependence upon the condensate density. The fitting parameter $\mathcal{C_{\infty}}$ of Eq.~\ref{Eq:Fit_tails} is equal to the Tan contact constant \cite{Tan2008, Combescot2009}, which, for a harmonically trapped gas, is found equal to $\mathcal{C}_{\text{LDA}}=(64 \pi^2 /7) a_{s}^2 N_{0} n_{0}$ in the LDA approximation \cite{SuppMaterial}. The experimental results are plotted in Fig.~\ref{Fig4} where the error bars reflect the uncertainty on $\mathcal{C_{\infty}}$ from the fit, as well as those on the calibration of $N_{0}$ and $n_{0}$ \cite{SuppMaterial}. The fitted contact constant $\mathcal{C_{\infty}}/N_{0}$ per condensed particle is found  proportional to $n_{0}$, as expected. The measured values of  $\mathcal{C}_{\infty}$, however, are about 6.5 times larger than the expected value $\mathcal{C}_{\text{LDA}}$. Note that in order to increase the density $n_{0}$ we increase the trapping frequency, which results in a decrease of the density of the central, dominant part of the distribution $n_{\infty}(k)$ measured after TOF. The observed proportionality between $\mathcal{C_{\infty}}/N_{0}$ and $n_{0}$ in spite of the variation of $n_{\infty}(k\simeq 0)$  rules out several possible spurious effects in the response of the MCP.


In conclusion, the measurement of the momentum distribution of a weakly interacting Bose gas released from a trap has allowed us to observe two components in the high momentum tails beyond the mean-field distribution. The first one is due to thermal depletion, and although some questions remain open, there are several observations which suggest associating the second one with the quantum depletion. The single-atom detection method of metastable Helium gases is also able to provide signals of atom-atom correlations in momentum, a feature we intend to use in future investigations of momentum-space signatures of many-body effects.

\begin{acknowledgments}
We thank E. Cornell, Z. Hadzibabic, K. M\"olmer, L. P. Pitaevskii, S. Stringari for thorough discussions about the interpretation of our results. We thank F. Nogrette for providing support on the detector and M. Mancini for reading the manuscript. We acknowledge financial support from the R\'egion Ile-de-France (DIM Daisy), the RTRA Triangle de la Physique, the European Research Council (Senior Grant Quantatop), the LabEx PALM (ANR-10-LABX-0039), the International Balzan Prize Foundation (2013 Prize for Quantum Information Processing and Communication awarded to A. Aspect), the Direction G\'en\'erale de l'Armement, the french National Research Agency (ANR 15-CE30-0017-04) and the Institut Francilien de Recherche sur les Atomes Froids.
\end{acknowledgments}



\begin{thebibliography}{21}
\bibitem{NoteTempHartree} We note that $T_{a}$ may slightly differ from the gas temperature $T$ in the regime $k_{B}T\simeq\mu$ due to the role of the Hartree-Fock potential induced by the condensate \cite{Dalfovo1999}.
\bibitem{NoteThermalContact} Contact interactions between thermal atoms lead to a thermal contribution to the $k^{-4}$-tails. The in-trap density of the thermal component is two orders of magnitude smaller than that of the condensate. The thermal contribution to the $k^{-4}$-tails should thus be negligible.
\end{thebibliography}

 \newpage

\section{Supplementary material}

\subsection{Production of gaseous condensates and detection after time-of-flight}

Our condensates are produced in the $2^3S_1$, $m_J=+1$ internal electronic state in a crossed dipole trap which is described in \cite{Bouton2015}. This metastable state has a lifetime of 8000 seconds. To measure the asymptotic momentum distribution $n_{\infty}({\bf k})$ of the gas, we switch off the optical trap in 2 $\mu$s, allowing the cloud to undergo expansion while falling under the influence of gravity. The Micro-Channel Plate (MCP) detector and electronics are described in detail in \cite{Nogrette2015}. The estimated detection efficiency is 25\%. The spatial resolution in the plane of the detector is measured to be equal to $\simeq 100$ $\mu$m and the time resolution along the vertical axis is 10 $\mu$s.

The electronic detector is positioned 55 cm below the trapped gas giving a TOF value of 330 ms. To ensure that the gas expansion is un-perturbed by residual magnetic field gradients present during the TOF, atoms are transferred to the magnetically insensitive $m_J=0$ state. This transfer is achieved with a radio-frequency (RF) which causes transitions between the magnetic sublevels whose energies have been split by a magnetic bias field with $\Delta E=E_{m_J=+1}-E_{m_J=0}\simeq h \times 10$ MHz. To transfer the atoms independently of their velocity, we use a RF sweep with central frequency 10 MHz and span of $\pm 500$ kHz, applied on the atoms after a  2 ms TOF. The RF sweep is 1 ms long at constant RF power. After the RF sweep, we apply a magnetic gradient to push the atoms remaining in the $m_J=\pm 1$ states away from the detector. The RF power, combined with the removal of  the $m_J=\pm 1$ states, allows us to control the flux of atoms ($m_J=0$) striking the detector. We typically operate between 15 and 45\% RF transfer efficiency (see \textbf{1D density profiles}). The scattering length in the $m_J=+1$ is $a_{s}^{m_J=+1}\simeq142 a_{0}$, where $a_{0}$ is the Borh radius. In the $m_J=0$ state, it has never been measured so we have inferred its value from the predictions of \cite{Leo2001} with the knowledge of $a_{s}^{m_J=+1}$ to extract the contribution of the quintet potential. We find $a_{s}^{m_J=0}\sim100 a_{0}$, much smaller than $a_{s}^{m_J=+1}$. Note that an experimental measurement might reveal corrections to this approximate value.

The MCP detector provides a three-dimensional histogram of atom numbers (see Fig.~\ref{Fig1}(b)). The position of an atom labelled with integer $j$ is given by two-dimensional spatial coordinates $(Y_{j},Z_{j})$ in the plane of the MCP and the time of arrival $t_{j}$. Similarly we note $(Y_{0},Z_{0},\bar{t}=t_{0})$ the coordinates of the center of the cloud whose time of arrival defines the TOF $\bar{t}$ used throughout this work. In the frame centered on the falling could, the position ${\bf r}$ of the atom is ${\bf r}=(g \bar{t}/2 t_{j} \times (\bar{t}^2-t_{j}^2), Y_{j}-Y_{0},Z_{j}-Z_{0})$, accounting for the acceleration $g$ of gravity. The use of the ballistic relation yields the asymptotic momentum distribution $n_\infty({\bf k}) = (\hbar \bar{t}/m)^3 \ n({\bf r}={\hbar \bf k} \bar{t}/m,\bar{t})$ (Eq.~\ref{Eq:asymp} in the main text). The momentum-space resolution is 0.03 $\mu m^{-1}$ along directions $y$ and $z$ (in the MCP plane), and 0.01 $\mu m^{-1}$ along $x$ (orthogonal to the MCP plane). The distributions studied here consist of the the average of roughly 1500 experimental shots. Individual shots have been re-centered to account for slight fluctuations in cloud center-of-mass after the TOF.  

The condensate atom number $N_{0}$ and density $n_{0}$ are calibrated by comparing the 3D density profiles on the MCP detector with the predictions of the scaling solution for our trap frequencies \cite{Castin1996, Kagan1996}. The uncertainty on these values is evaluated to be 20\%.

\subsection{1D density profiles}
 
\begin{figure*}[ht!]
\begin{center}
\includegraphics[width=1.8\columnwidth]{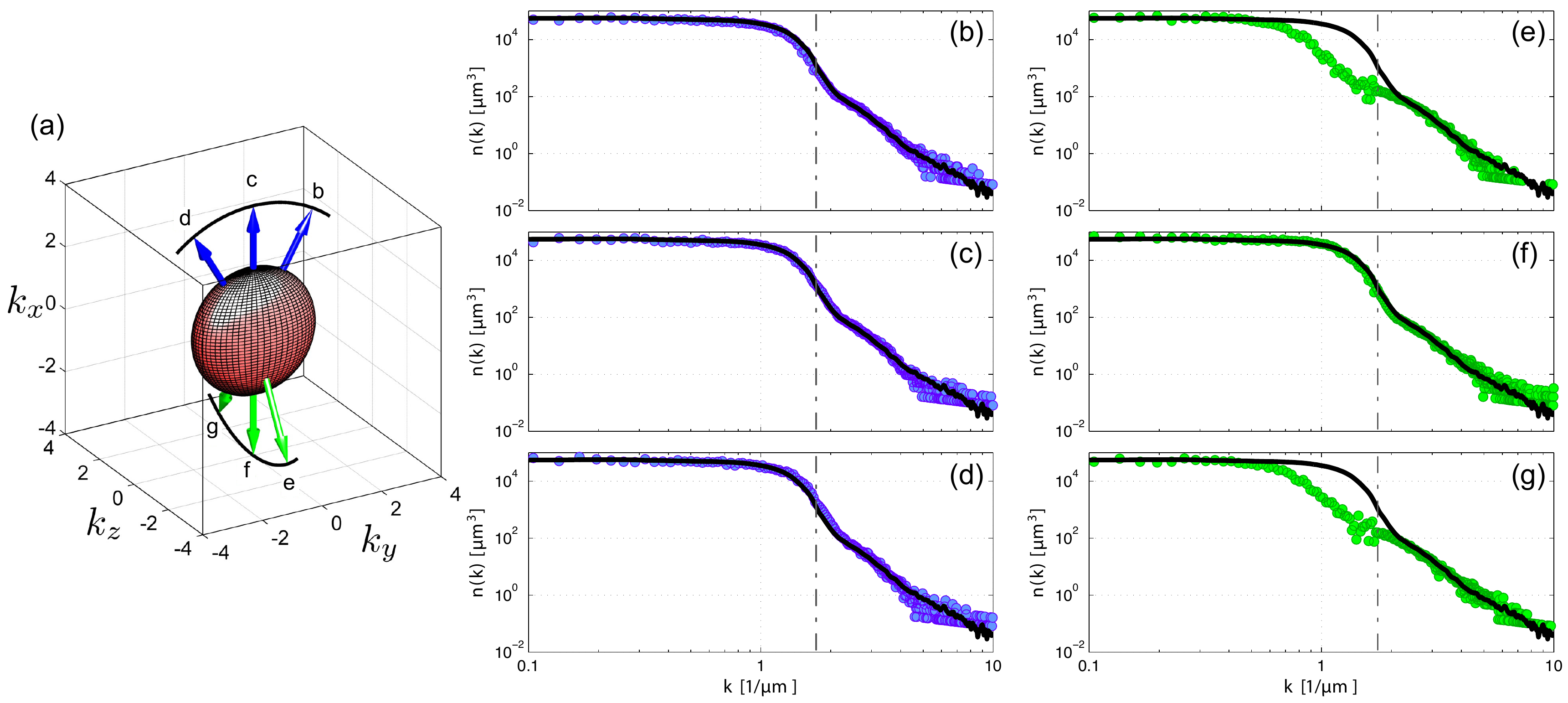}
\caption{Observed momentum distributions along various directions, showing the anisotropic condensate distribution and the isotropic components associated with the thermal and the quantum depletion. The vertical dot-dashed line delimits the condensate (low momenta $k$) and the condensate depletion.  {\bf (a)} Illustration of the anisotropic momentum distribution of the condensate in the far-field regime (pancake-shaped distribution). The arrows represent the directions along which we plot the 1D momentum profiles with a small angular average ($\pm 10^\circ$ degrees) in the panels (b)-(g). {\bf (b)-(d)} Blue dots are the measured 1D momentum profiles in the $k_{z}=0$ plane. {\bf (e)-(g)} Green dots are the measured 1D momentum profiles in the $k_{y}=0$ plane. In each subplot, the solid line is the radial 1D momentum profile plotted in Fig.~\ref{Fig2} of the main text (obtained from a large angular average, see above). \label{FigSup1}}
\end{center}
\end{figure*}

From the three-dimensional asymptotic momentum density $n_{\infty}({\bf k})$, one-dimensional profiles are generated for the radial and longitudinal cloud directions. Longitudinal profiles are cuts along the $z$ direction (at $k_{x}\simeq k_{y} \simeq 0$) with a small integration along the two transverse $x$ and $y$ directions (see below). Radial profiles in the $x-y$ plane are generated from an angular average over azimuthal angles $|\phi|>45^\circ$, where $\phi=\arctan(k_{x}/k_{y})$ (at $k_{z}\simeq 0$). This choice of azimuthal angles avoids a defect located on the surface of the micro-channel plate.  

Since the experimental signals contain both a high-density condensate and low-density tails with a 4 decade separation in scales, it is necessary to divide the measurement of the density profile into two steps. To measure the condensate momentum components we first use a low RF transfer efficiency (typically 15\%). In this low particle flux regime the MCP is far from electronic saturation, but the low-density tails are hard to detect. Secondly to measure the high-momentum tails, we use a high RF transfer efficiency (typically 45\%).  At high flux of detected particles, the dense condensate locally saturates the MCP, while the low-density wings remain unperturbed. We have verified that the two runs provide identical density profiles at intermediate $k$ where local saturation is not a problem for the high-flux runs and there is sufficient signal in the low-flux runs. The 1D profile for the condensate (low-flux data) has a transverse integration of $\pm$ 0.1 $\mu m^{-1}$. The profile for the tails (high-flux data) has transverse integration of $\pm$ 0.8 $\mu m^{-1}$.  The two-step measurement and transverse integration ensures sufficient signal in the high-momentum tails, while accurately capturing the condensate profile at low-momenta. 

In order to illustrate the observed symmetry in region {\bf II} and {\bf III}, we plot measured 1D profiles with a small angular average on $\phi$ ($\pm 10^\circ$ degrees) along different directions separated by 30$^\circ$degrees in Fig.~\ref{FigSup1}. The radial 1D profile shown in Fig.~\ref{Fig2} (obtained from a large angular average) is reported as a black line in each subplot. The anisotropy of the condensate distribution appears clearly while the momentum profiles in region {\bf II} and {\bf III} have a spherical symmetry.

\subsection{Mean-field momentum distribution of an interacting gas from time-of-flight}

The condensate and its expansion is known to be well modeled by a mean-field interaction \cite{Dalfovo1999}. To verify that the observed tails (regions {\bf II} and {\bf III} in Fig.~\ref{Fig2}(b)) are not an artifact of the condensate, we compare our data to the complete 3D mean-field solution. The mean-field Gross-Pitaevskii (GP) solution beyond the Thomas-Fermi approximation may lead to the appearance of additional momentum components after TOF. These would result from slight modifications of the Thomas-Fermi real-space density occurring on the length scale $(a^4_{ho}/R)^{1/3}$, which fixes the characteristic thickness of the boundary \cite{Dalfovo1999} (here $a_{ho}$ and $R$ are, respectively, the oscillator length and the Thomas-Fermi radius and, for simplicity, we have assumed isotropic trapping). 

\begin{figure}[hb]
\begin{center}
\includegraphics[width=0.8\columnwidth]{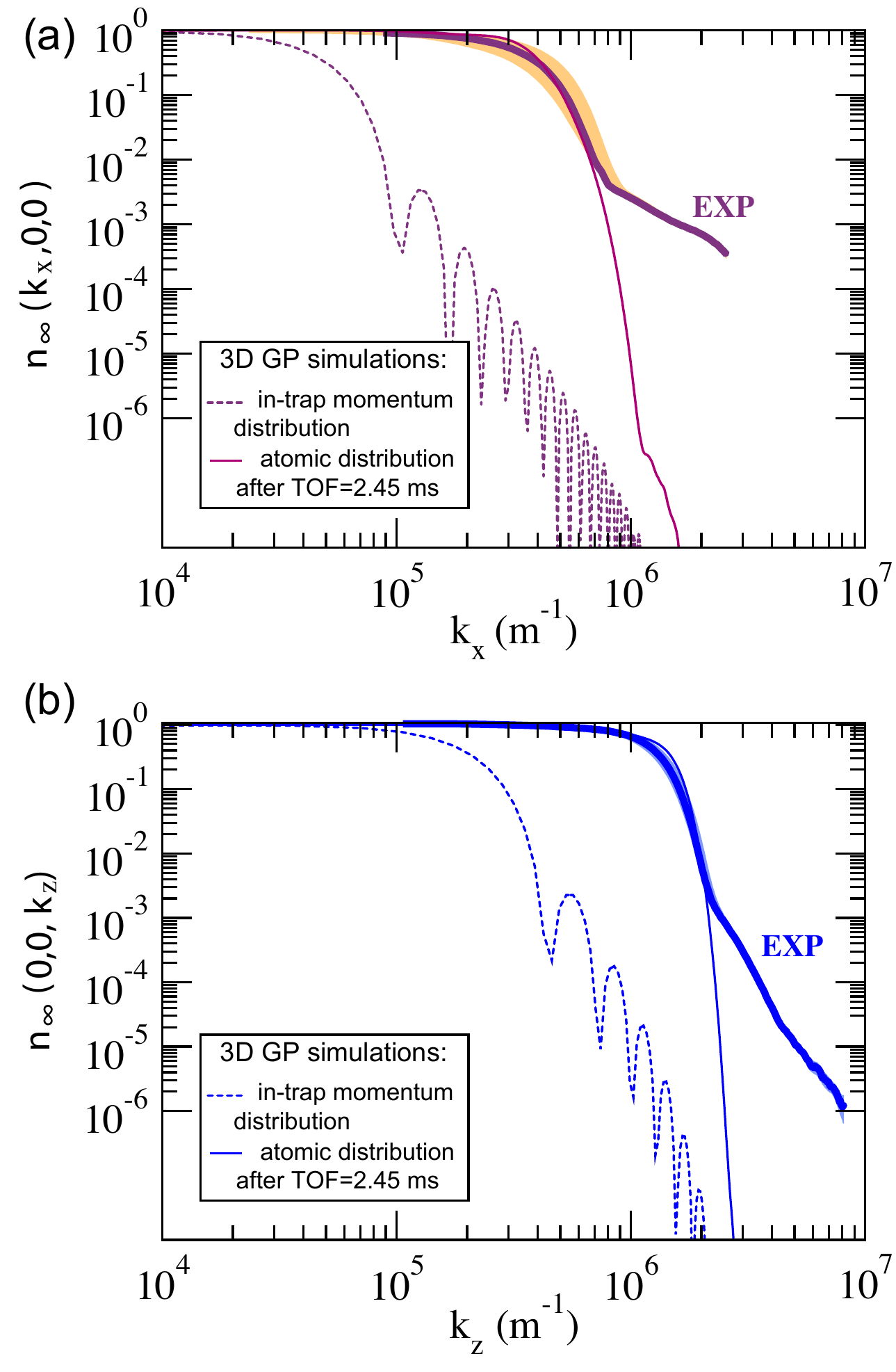}
\caption{Simulation of the expansion dynamics of a 3D Bose-Einstein condensate with the mean-field Gross-Pitaevskii equation. The parameters describing the condensate in the numerics (including atom number and trapping frequencies) are those of the experiment. The GP simulation is compared to the experimental data along the longitudinal {\bf (a)} and radial {\bf (b)} direction. It shows an excellent agreement in the region {\bf I} associated with the condensate but does not reproduce the experimental tails. Numerical results and experimental distributions have been normalized to $n_{\infty}(0,0,0)=1$. \label{FigSup2}}
\end{center}
\end{figure}

The ground state is obtained by numerical simulation of the Gross-Pitaevskii equation in imaginary time, and the expansion dynamics are performed through real time propagation. The system size used in the calculations limits the simulated TOF. However, the distribution after 2.45 ms is observed to converge, indicating the complete conversion of mean-field interaction energy into kinetic energy. The numerical results for the distribution after a TOF of 2.45 ms are presented in Fig.~\ref{FigSup2}, using the ballistic relation as defined in Eq.~\ref{Eq:asymp}. The simulations clearly show that the mean-field Gross-Pitaevskii ground state is unable to reproduce the tails observed in the experiment.  \\

\subsection{Quantum depletion with local density approximation at $T=0$}

In uniform systems Bogoliubov's approach yields a population of momentum state equal to 
\begin{equation}
n(k)=|v_k|^2=\frac{\hbar^2k^2/2m+mc^2}{2\epsilon(k)}-\frac{1}{2}
\end{equation}
where $\epsilon(k)=\left[\hbar^2k^2c^2+(\hbar^2k^2/2m)^2 \right]^{1/2}$ is the Bogoliubov excitation spectrum. For a harmoncially trapped condensate in the Thomas-Fermi limit, the local speed of sound is $c({\bf r})=c_0\sqrt{1-(x/R_x)-(y/R_y)^2-(z/R_z)^2}$ with $c_{0}=\sqrt{g_{s} n_{0}/m}$ the speed of sound at the trap center and $g_{s}=4 \pi \hbar^2 a_{s}/m$. In the LDA one finds

\begin{widetext}
\begin{eqnarray}
n(\vec{k})&=&\frac{1}{(2\pi)^3}\int |v_{\vec{k}}(\vec{r})|^2 d\vec{r}\\
&=&\frac{R_xR_yR_z}{(2\pi)^2}\left[ -\frac{13}{48}-\frac{5k^2\xi^2}{32}+\left( \frac{4+12k^2\xi^2+5k^4\xi^4}{32\sqrt{2}k\xi} \right)\arctan\left( \frac{\sqrt{2}}{k\xi} \right)  \right] \\
& \underset{[k\xi \gg 1]}{\simeq}&
\frac{R_xR_yR_z}{105\pi^2}\frac{1}{k^4\xi^4}
\end{eqnarray}
\end{widetext}

\noindent where $\xi=\hbar/\sqrt{2}mc_{0}$ is the healing length.

Within the Bogoliubov approach in the LDA approximation, the Tan contact constant, defined as $\mathcal{C}/(2 \pi)^3=\text{lim}_{k\to \infty} n(k)k^4$, is equal to $\mathcal{C}_{\text{LDA}}=(64 \pi^2 /7)  a_{s}^2 N_{0} n_{0}$, with $a_{s}$ the s-wave scattering length and $N_{0}$ ({\it resp.} $n_{0}$) the condensate atom number ({\it resp.} the condensate density). For metastable Helium, the calculation gives  $(\mathcal{C}_{\text{LDA}}/N_{0})/n_{0} \simeq 5.08 \times 10^{-15}$ m$^{-2}$.

\subsection{Controlled heating sequence}

To increase the temperature of our ultracold gas, we perform a controlled heating sequence using a 3D optical lattice.  This sequence involves the adiabatic transfer of the gas from the optical dipole trap to the lattice in 30 ms, followed by a series of non-adiabatic lattice pulses of duration 0.5 ms during which the amplitude of the lattice is set to zero. The gas is then held in the optical lattice for 100 ms during which time it rethermalizes. Finally, we transfer the atom cloud adiabatically back to the initial optical dipole trap in 30 ms. Increasing the lattice depth while keeping the same sequence increases the final temperature of the gas at constant atom number and trapping frequencies.

\end{document}